\documentclass[10pt,twocolumn]{article}
\usepackage{wrapfig}
\usepackage{graphicx}
\usepackage{amsmath,amssymb}

\title{\textbf{On the response of a system with bound states of particles to the
perturbation by the external electromagnetic field}}

\author{Yuriy V. Slyusarenko and Andrey G. Sotnikov
    \vspace{3mm}
    \\
\emph{\small Akhiezer Institute for Theoretical Physics, NSC KIPT,
                1~Akademicheskaya str., 61108 Kharkiv, Ukraine}}
\date{}

\begin{document}
\twocolumn[\maketitle \small{
    The response of the system, consisting of two kinds of
    opposite-charged fermions and their bound states
    (hydrogen-like atoms), to the perturbation by the external
    electromagnetic field in low particle kinetic energies region
    is studied. Expressions for Green functions
    that describe the system response to the external
    electromagnetic field and take into account the presence of
    particle bound states (atoms) are found. Macroscopic parameters
    of the system, such as conductivity, permittivity and
    magnetic permeability in terms of these Green
    functions are introduced. As an example, the perturbation of the ideal
    hydrogen-like plasma by the external electromagnetic field in
    low temperature region is considered.
    Such approach also enables to study the propagation properties
    of the signal, tuned up to the transition between two hyperfine ground
    state levels of alkali atoms that are considered in Bose-Einstein
    condensation (BEC) state. It is shown that the signal can propagate
    in such system with rather small energy loss.
    Such fact allows to introduce the group
    velocity concept and to study the slowing down conditions for
    the microwave signal that propagates in BEC.
    \vspace{3mm}
    \vspace{4mm}
    \\}]
\section{Introduction}
In the process of describing a behavior of many-particle systems a
class of problems appear, that concerned with the system response to
the perturbing action of the external, in particular,
electromagnetic field. Widespread approach to solving such kind of
problems is based on using the Green functions formalism (see in
that case e.g. \cite{ref:01}).

As well known, the most convenient method for physical processes
describing in quantum many-particle theory is the second
quantization method. Because of that in framework of the second
quantization it is the most simple to formulate an approach for
describing the system response to the perturbation by the external
field, that is based on using Green functions. But if we try to
realize such an approach, we can meet an essential difficulty,
connected with the possibility of the particle bound states
existence.

Really, the key role of the second quantization method consists in
the introduction of creation and annihilation operators of
particles in a certain quantum state. The operators of physical
quantities are constructed in terms of creation and annihilation
operators. Such a description of quantum many-particle systems
implies the particles to be elementary (not consisting of other
particles). Moreover, it is absolutely accurate despite of the
possible existence of compound particles. Since the interactions
between particles may lead to the formation of bound states, the
standard second quantization method becomes too cumbersome. For
this reason the construction of an approximate quantum--mechanical
theory for many--particle systems consisting of elementary
particles and their bound states represents an actual problem. In
this theory it is necessary to introduce the creation and
annihilation operators of bound states as the operators of
elementary objects (not compound). In addition it must preserve
the required information concerning internal degrees of freedom
for bound states.

Such an approach is realized in \cite{ref:02}. In this work the
possibility of constructing such theory is demonstrated for a
system, that consists of two kinds of fermions, assuming that bound
states (atoms or molecules) are formed by particles of two different
kinds. The choice of such model is not dictated by the principle
difficulties but by the desire to simplify calculations and obtain
the visual results. In framework of this model a method of
constructing the creation and annihilation operators of the bound
state as a compound object is given. The substantiation of the
conversion from the description of atoms as compound objects to
elementary objects with the ordinary creation and annihilation
Bose-operators is given. Such substantiation is considered in
low-energy approximation in which the binding energy of a compound
particle is much greater than its kinetic energy. In terms of the
creation and annihilation operators of fermions and bosons (as
elementary objects) a scheme for physical quantities operators
construction is formulated. Explicit expressions for the operators
of principal physical quantities, such as density and charge
density, momentum and current density, system Hamiltonian, are
found. The Maxwell--Lorentz system of equations, describing the
interaction between electromagnetic field and matter, that may
consist also of neutral ''atoms'' (low-energy quantum electrodynamic
equations), is found.

In the present work we use this system of equations to study in the
framework of Green functions formalism the response of the system
with bound states to the perturbation by the external
electromagnetic field. An essentially new moment in these
considerations is the next circumstance. When we describe the system
response to the perturbing action of the external electromagnetic
field the approximate formulation of the second quantization method
proposed in \cite{ref:02} gives us the possibility to account for
the neutral bound states in sufficiently simple way.

\section{Quantum electrodynamic equations
for the low-temperature hydrogen-like plasma}

The studied quantum-electrodynamic system, consisting of fermions of
two different kinds and their bound states, in low-energy region, in
fact, can be considered as a low-temperature hydrogen-like plasma.
Before we turn to the description of such plasma response to the
external electromagnetic field, let us obtain the main equations
that describe an evolution of such system. Taking into account the
interaction between radiation and matter the system's Hamiltonian
$\hat{\mathcal H}(t)$, according to \cite{ref:02}, can be written as
\begin{equation}\label{eq:01}                               %%%% ---- %%%% eq 1
    \hat{\mathcal H}(t)
    =\hat{\mathcal H}_0
    +\hat{\mathcal H}_{int}
    +\hat{V}(t),\quad
    \hat{\mathcal H}_0
    =\hat{\mathcal H}_{f}
    +\hat{\mathcal H}_{p},
\end{equation}
where
\begin{equation}\label{eq:02}                               %%%% ---- %%%% eq 2
    \hat{\mathcal H}_{f}
    =\sum_{\textbf{k},\lambda}\omega_k
    {\hat C}^{+}_{\textbf{k}\lambda}
    {\hat C}_{\textbf{k}\lambda}
\end{equation}
is the Hamiltonian for free photons ($\omega_k$ is the frequency of
photon with wave number~$k$, ${\hat C}^{+}_{\textbf{k}\lambda}$,
${\hat C}_{\textbf{k}\lambda}$ are the creation and annihilation
operators of photon with wave number~$k$ and polarization
$\lambda$).

The value $\hat{\mathcal H}_{p}$ in the equation \eqref{eq:01} is
the Hamiltonian for free particles (free fermions and their bound
states)
\begin{equation}\label{eq:03}                               %%%% ---- %%%% eq 3
\begin{split}
    \hat{\mathcal H}_{p}
    =&\sum_{j=1}^{2}{1\over 2m_{j}}
    \int d\textbf{x}\dfrac{\partial{\hat\chi}_{j}^{+}
    (\textbf{x})}{\partial\textbf{x}}
    \dfrac{\partial{\hat\chi}_{j}(\textbf{x})}
    {\partial\textbf{x}}
    \\
    &+\sum_{\alpha}\int d\textbf{X}
    \left\{
    \dfrac{1}{2M}
    \dfrac{\partial{\hat\eta}_{\alpha}^{+}(\textbf{X})}
    {\partial\textbf{X}}
    \dfrac{\partial{\hat\eta}_{\alpha}
    (\textbf{X})}{\partial\textbf{X}}\right.
    \\
    &\left.+\varepsilon_{\alpha}{\hat\eta}_{\alpha}^{+}
    (\textbf{X}){\hat\eta}_{\alpha}(\textbf{X})
    \right\},
    \quad
    M=m_{1}+m_{2},
\end{split}
\end{equation}
where ${\hat\chi}_{j}^{+}(\textbf{x})$,
${\hat\chi}_{j}(\textbf{x})$ ($j=1,2$) are the creation and
annihilation operators of a free fermion of $j$ kind and mass
$m_{j}$ at the point $\textbf{x}$;
${\hat\eta}_{\alpha}^{+}(\textbf{X})$, ${\hat\eta}_{\alpha}
(\textbf{X})$ are the creation and annihilation operators of bound
states of two different fermions (''hydrogen-like atoms'') with
the quantum numbers $\alpha$ at the point $\textbf{X}$;
$\varepsilon_{\alpha}$ is the energy of an atom at the level with
the quantum numbers $\alpha$.

The Hamiltonian $\hat{\mathcal H}_{int}$ in the equation
\eqref{eq:01} describes the interaction between particles
\begin{equation}\label{eq:04}                               %%%% ---- %%%% eq 4
    \hat{\mathcal H}_{int}
    =\hat{\mathcal H}_{int}^{1}
    +\hat{\mathcal H}_{int}^{2}
    +\hat{\mathcal H}_{int}^{3},
\end{equation}
where
\begin{equation}\label{eq:05}                               %%%% ---- %%%% eq 5
\begin{split}
    &\hat{\mathcal H}_{int}^{1}
    =\int d\textbf{x}_{1}d\textbf{x}_{2}d\textbf{y}
    {\hat\varphi}^{+}(\textbf{x}_{2},\textbf{y})
    {\hat\varphi}(\textbf{x}_{2},\textbf{y})
    \\
    &\times
    \{(\nu_{11}(\textbf{x}_{1}-\textbf{x}_{2})
    +\nu_{21}(\textbf{x}_{1}-\textbf{y}))
    {\hat\chi}^{+}_{1}(\textbf{x}_{1})
    {\hat\chi}_{1}(\textbf{x}_{1})
    \\
    &+(\nu_{22}(\textbf{x}_{1}-\textbf{y})
    +\nu_{12}(\textbf{x}_{1}-\textbf{x}_{2}))
    {\hat\chi}^{+}_{2}(\textbf{x}_{1})
    {\hat\chi}_{2}(\textbf{x}_{1})\},
\end{split}
\end{equation}
\begin{equation}\label{eq:06}                               %%%% ---- %%%% eq 6
\begin{split}
    \hat{\mathcal H}_{int}^{2}
    ={1\over 2}\int d&\textbf{x}_{1}d\textbf{x}_{2}
    d\textbf{y}_{1}d\textbf{y}_{2}
    {\hat\varphi}^{+}(\textbf{x}_{1},\textbf{y}_{1})
    \\
    &\times
    {\hat\varphi}^{+}(\textbf{x}_{2},\textbf{y}_{2})
    {\hat\varphi}(\textbf{x}_{2},\textbf{y}_{2})
    {\hat\varphi}(\textbf{x}_{1},\textbf{y}_{1})
    \\
    &\times
    \{\nu_{11}(\textbf{x}_{1}-\textbf{x}_{2})
    +\nu_{22}(\textbf{y}_{1}-\textbf{y}_{2})
    \\
    &+\nu_{12}(\textbf{x}_{1}-\textbf{y}_{2})
    +\nu_{21}(\textbf{y}_{1}-\textbf{x}_{2})\},
\end{split}
\end{equation}
\begin{equation}\label{eq:07}                               %%%% ---- %%%% eq 7
\begin{split}
    &\hat{\mathcal H}_{int}^{3}
    ={1\over 2}\int d\textbf{x}_{1}d\textbf{x}_{2}
    \\
    &\times
    \{\nu_{11}(\textbf{x}_{1}-\textbf{x}_{2})
    {\hat\chi}^{+}_{1}(\textbf{x}_{1})
    {\hat\chi}^{+}_{1}(\textbf{x}_{2})
    {\hat\chi}_{1}(\textbf{x}_{2})
    {\hat\chi}_{1}(\textbf{x}_{1})
    \\
    &+\nu_{22}(\textbf{x}_{1}-\textbf{x}_{2})
    {\hat\chi}^{+}_{2}(\textbf{x}_{1})
    {\hat\chi}^{+}_{2}(\textbf{x}_{2})
    {\hat\chi}_{2}(\textbf{x}_{2})
    {\hat\chi}_{2}(\textbf{x}_{1})
    \\
    &+2\nu_{12}(\textbf{x}_{1}-\textbf{x}_{2})
    {\hat\chi}^{+}_{1}(\textbf{x}_{1})
    {\hat\chi}^{+}_{2}(\textbf{x}_{2})
    {\hat\chi}_{2}(\textbf{x}_{2})
    {\hat\chi}_{1}(\textbf{x}_{1})\}.
\end{split}
\end{equation}

In equations \eqref{eq:05}--\eqref{eq:07} the operators
${\hat\varphi}^{+} (\textbf{x}_{1},\textbf{x}_{2})$,
${\hat\varphi}(\textbf{x}_{1}, \textbf{x}_{2})$ are related to the
creation ${\hat\eta}_{\alpha}^{+}(\textbf{X})$ and annihilation
operators ${\hat\eta}_{\alpha}(\textbf{X})$ of atoms in quantum
state $\alpha$ by expressions
\begin{equation}\label{eq:08}                               %%%% ---- %%%% eq 8
\begin{split}
    &{\hat\varphi}^{+}(\textbf{x}_{1},\textbf{x}_{2})
    =\sum_{\alpha}\varphi_{\alpha}^{*}(\textbf{x})
    {\hat\eta}_{\alpha}^{+}(\textbf{X}),
    \\
    &{\hat\varphi}(\textbf{x}_{1},\textbf{x}_{2})
    =\sum_{\alpha}\varphi_{\alpha}(\textbf{x})
    {\hat\eta}_{\alpha}(\textbf{X}),
    \\
    &\textbf{x}=\textbf{x}_{1}-\textbf{x}_{2},
    \quad \textbf{X}=\frac{m_{1}\textbf{x}_{1}
    +m_{2}\textbf{x}_{2}}{m_{1}+m_{2}}~,
\end{split}
\end{equation}
where $\varphi_{\alpha}(\textbf{x})$ is the wave function of the
bound state and $\nu_{ij}(\textbf{x} -\textbf{y}),\ i,j=1,2$ is the
potential energy of Coulomb interaction
\begin{equation}\label{eq:09}                               %%%% ---- %%%% eq 9
    \nu_{ij}(\textbf{x}-\textbf{y})
    ={e_{i}e_{j}\over\vert\textbf{x}-\textbf{y}\vert}
\end{equation}
($e_{i}$ is the electric charge of a fermion of kind $i$).

In that way, the Hamiltonian $\hat{\mathcal H}_{int}^{1}$
corresponds to scattering of particles of the first and second kinds
by bound states, the Hamiltonian $\hat{\mathcal H}_{int}^{2}$
corresponds to scattering of bound states by bound states, the
Hamiltonian $\hat{\mathcal H}_{int}^{3}$ corresponds to scattering
of particles of the first and second kinds by particles of the same
kinds.

And, finally, the operator $\hat{V}(t)$ in \eqref{eq:01}
represents the Hamiltonian that describes the interaction of
particles with the electromagnetic field
\begin{equation}\label{eq:10}                               %%%% ---- %%%% eq 10
\begin{split}
    &\hat{V}(t)
    =-\dfrac{1}{c}\int d\textbf{x}\hat{\textbf{A}}(\textbf{x},t)
    \hat{\textbf{J}}(\textbf{x},t)
    \\
    &-\dfrac{1}{2c^{2}}
    \int d\textbf{x}\hat{\textbf{A}}^{2}(\textbf{x},t)
    \sum\limits_{i=1}^{2}\dfrac{e_{i}}{m_{i}}\hat{\sigma}_{i}
    (\textbf{x})
    \\
    &+\int d\textbf{x}\varphi^{(e)}(\textbf{x},t)
    \hat{\sigma}(\textbf{x}),\quad
    \hat{\sigma}(\textbf{x})
    =\sum_{i=1}^{2}\hat{\sigma}_{i}(\textbf{x}).
\end{split}
\end{equation}
In this expression we have taken into account an interaction of
particles with the external electromagnetic field
$\textbf{A}^{(e)}(\textbf{x},t)$, $\varphi^{(e)}({\bf x},t)$
($\varphi^{(e)}(\textbf{x},t)$ is the scalar potential of the
external electromagnetic field) and the quantum electromagnetic
field, that is described by the potential
$\hat{\textbf{a}}(\textbf{x})$ (Coulomb's gauge):
\begin{equation}\label{eq:11}                               %%%% ---- %%%% eq 11
    \hat{\textbf{A}}(\textbf{x},t)
    =\hat{\textbf{a}}(\textbf{x})
    +\textbf{A}^{(e)}(\textbf{x},t),
\end{equation}
where $\textbf{A}^{(e)}(\textbf{x},t)$ is the vector potential of
the external electromagnetic field and
$\hat{\textbf{a}}(\textbf{x})$ is the quantum electromagnetic field
operator, that is defined by the expression
\begin{equation*}
    \hat{\textbf{a}}(\textbf{x})
    =\sum_{\textbf{k}}\sum_{\lambda=1}^{2}
    \left({2\pi\over V\omega_{\textbf{k}}}
    \right)^{1/2}\left(\textbf{e}_{\textbf{k}\lambda}
    {\hat C}_{\textbf{k}\lambda}
    e^{i\textbf{k}\textbf{x}}+h.c.\right)
\end{equation*}
($V$ is the system volume, $\textbf{e}_{\textbf{k}\lambda}$ is the
photon polarization vector).

The charge density operators $\hat{\sigma}_{i}(\textbf{x})$ for
particles of $i$ kind (see \eqref{eq:10}) are connected with the
density operators ${\hat\rho}_{i}(\textbf{x})$ (see \cite{ref:02})
\begin{equation}\label{eq:12}                               %%%% ---- %%%% eq 12
\begin{split}
    \hat{\sigma}_{i}(\textbf{x})
    &=e_{i}{\hat\rho}_{i}(\textbf{x}),
    \\
    \hat{\rho}_{1}(\textbf{x})
    &=\hat{\chi}_{1}^{+}(\textbf{x})
    \hat{\chi}_{1}(\textbf{x})
    +\int d\textbf{y}\int d\textbf{Y}
    \\
    &\times
    \delta(\textbf{x}-\textbf{Y}
    -\dfrac{m_{2}}{M}\textbf{y})
    \hat{\varphi}^{+}(\textbf{y},\textbf{Y})
    \hat{\varphi}(\textbf{y},\textbf{Y}),
    \\
    \hat{\rho}_{2}\textbf{x})
    &=\hat{\chi}_{2}^{+}(\textbf{x})
    \hat{\chi}_{2}(\textbf{x})
    +\int d\textbf{y}\int d\textbf{Y}
    \\
    &\times
    \delta(\textbf{x}-\textbf{Y}
    -\dfrac{m_{1}}{M}\textbf{y})
    \hat{\varphi}^{+}(\textbf{y},\textbf{Y})
    \hat{\varphi}(\textbf{y},\textbf{Y}),
\end{split}
\end{equation}
and, as easy to see, in the equation \eqref{eq:12} we have also
taken into account a contribution that had been made by charged
particles, that are represented in the bound states (see
\eqref{eq:08}). The current density operator
$\hat{\textbf{J}}(\textbf{x},t)$ in the formula \eqref{eq:10} can be
also expressed in terms of the creation and annihilation operators
\begin{equation}\label{eq:13}                               %%%% ---- %%%% eq 13
\begin{split}
    \hat{\textbf{J}}(\textbf{x},t)
    =-\hat{\textbf{A}}(\textbf{x},t)
    \sum\limits_{i=1}^{2}\dfrac{e_{i}}{m_{i}}
    \hat{\sigma}_{i}(\textbf{x})
    +\hat{\textbf{j}}(\textbf{x}),
    \\
    \hat{\textbf{j}}(\textbf{x})=
    \sum\limits_{i=1}^{2}\dfrac{e_{i}}{m_{i}}
    \hat{\boldsymbol{\pi}}_{i}(\textbf{x}),
\end{split}
\end{equation}
where the momentum density operators $\hat{\boldsymbol{\pi}}_{i}
(\textbf{x})$ are defined by expressions
\begin{equation}\label{eq:14}                               %%%% ---- %%%% eq 14
\begin{split}
    &\hat{\boldsymbol{\pi}}_{1}(\textbf{x})=
    -\dfrac{i}{2}\left(
    \hat{\chi}_{1}^{+}(\textbf{x})
    \dfrac{\partial\hat{\chi}_{1}(\textbf{x})}
    {\partial\textbf{x}}
    -\dfrac{\partial\hat{\chi}_{1}^{+}
    (\textbf{x})}{\partial\textbf{x}}
    \hat{\chi}_{1}(\textbf{x})\right)
    \\
    &-\dfrac{i}{2}\int d\textbf{y}\int d\textbf{Y}
    \delta(\textbf{x}-\textbf{Y}
    -\dfrac{m_{2}}{M}\textbf{y})
    \\
    &\times\left[
    \hat{\varphi}^{+}(\textbf{y},\textbf{Y})
    \dfrac{\partial\hat{\varphi}(\textbf{y},\textbf{Y})}
    {\partial\textbf{y}}
    -\dfrac{\partial\hat{\varphi}^{+}
    (\textbf{y},\textbf{Y})}{\partial\textbf{y}}
    \hat{\varphi}(\textbf{y},\textbf{Y})\right.
    \\
    &\qquad +
    \dfrac{m_{1}}{M}\left(
    \hat{\varphi}^{+}(\textbf{y},\textbf{Y})
    \dfrac{\partial\hat{\varphi}
    (\textbf{y},\textbf{Y})}{\partial\textbf{Y}}
    \right.
    \\
    &\left.\left.
    \qquad\qquad-\dfrac{\partial\hat{\varphi}^{+}
    (\textbf{y},\textbf{Y})}{\partial\textbf{Y}}
    \hat{\varphi}(\textbf{y},\textbf{Y})
    \right)\right],
    \\
    &\hat{\boldsymbol{\pi}}_{2}(\textbf{x})=
    -\dfrac{i}{2}\left(
    \hat{\chi}_{2}^{+}(\textbf{x})
    \dfrac{\partial\hat{\chi}_{2}
    (\textbf{x})}{\partial\textbf{x}}
    -\dfrac{\partial\hat{\chi}_{2}^{+}
    (\textbf{x})}{\partial\textbf{x}}
    \hat{\chi}_{2}(\textbf{x})\right)
    \\
    &-\dfrac{i}{2}\int d\textbf{y}\int d\textbf{Y}
    \delta(\textbf{x}-\textbf{Y}
    +\dfrac{m_{1}}{M}\textbf{y})
    \\
    &\times\left[
    -\hat{\varphi}^{+}(\textbf{y},\textbf{Y})
    \dfrac{\partial\hat{\varphi}
    (\textbf{y},\textbf{Y})}{\partial\textbf{y}}
    +\dfrac{\partial\hat{\varphi}^{+}
    (\textbf{y},\textbf{Y})}{\partial\textbf{y}}
    \hat{\varphi}(\textbf{y},\textbf{Y})\right.
    \\
    &\qquad+
    \dfrac{m_{2}}{M}\left(
    \hat{\varphi}^{+}(\textbf{y},\textbf{Y})
    \dfrac{\partial\hat{\varphi}
    (\textbf{y},\textbf{Y})}{\partial\textbf{Y}}
    \right.
    \\
    &\left.\left.
    \qquad\qquad-\dfrac{\partial\hat{\varphi}^{+}
    (\textbf{y},\textbf{Y})}{\partial\textbf{Y}}
    \hat{\varphi}(\textbf{y},\textbf{Y})
    \right)\right].
\end{split}
\end{equation}

Using formulas \eqref{eq:12}--\eqref{eq:14} we can write the
expressions for the current and charge density operators in more
suitable way:
\begin{equation}\label{eq:15}                              %%%% ---- %%%% Eq. 15
\begin{split}
    &\hat{\sigma}(\textbf{x})=\sum_{a}
    \hat{\sigma}_{a}(\textbf{x}),
    \\
    &\hat{\textbf{j}}(\textbf{x})=\sum_{a}\hat{\textbf{j}}_{a}(\textbf{x}),
    \quad a=0,1,2,
\end{split}
\end{equation}
where
\begin{equation}\label{eq:16}                              %%%% ---- %%%% Eq. 16
\begin{split}
    &\hat{\sigma}_{i}(\textbf{x})=
    e_{i}\hat{\chi}_{i}^{+}(\textbf{x})
    \hat{\chi}_{i}(\textbf{x}), \quad i=1,2,
    \\
    &\hat{\textbf{j}}_{i}(\textbf{x})=
    -\dfrac{ie_{i}}{2m_{i}}\left(
    \hat{\chi}_{i}^{+}(\textbf{x})
    \dfrac{\partial\hat{\chi}_{i}(\textbf{x})}
    {\partial\textbf{x}}
    -\dfrac{\partial\hat{\chi}_{i}^{+}(\textbf{x})}
    {\partial\textbf{x}}
    \hat{\chi}_{i}(\textbf{x})\right),
    \\
    &\hat{\sigma}_{0}(\textbf{x})=
    \int d\textbf{y}\int d\textbf{Y}
    \left[
    e_{1}\delta(\textbf{x}-\textbf{Y}
    -\dfrac{m_{2}}{M}\textbf{y})\right.
    \\
    &\left.\qquad
    +e_{2}\delta(\textbf{x}-\textbf{Y}
    -\dfrac{m_{1}}{M}\textbf{y})
    \right]
    \hat{\varphi}^{+}(\textbf{y},\textbf{Y})
    \hat{\varphi}(\textbf{y},\textbf{Y}),
    \\
    &\hat{\textbf{j}}_{0}(\textbf{x})=
    -\dfrac{i}{2}\int d\textbf{y}\int d\textbf{Y}
    \\
    &\times\left[
    \dfrac{e_{1}}{m_{1}}\delta(\textbf{x}-\textbf{Y}-
    \dfrac{m_{2}}{M}\textbf{y})
    -\dfrac{e_{2}}{m_{2}}\delta(\textbf{x}-\textbf{Y}-
    \dfrac{m_{1}}{M}\textbf{y})
    \right]
    \\
    &\times\left(
    \hat{\varphi}^{+}(\textbf{y},\textbf{Y})
    \dfrac{\partial\hat{\varphi}
    (\textbf{y},\textbf{Y})}{\partial\textbf{y}}
    -\dfrac{\partial\hat{\varphi}^{+}
    (\textbf{y},\textbf{Y})}{\partial\textbf{y}}
    \hat{\varphi}(\textbf{y},\textbf{Y})\right)
    \\
    &\qquad\quad-\dfrac{i}{2M}\int d\textbf{y}
    \int d\textbf{Y}
    \\
    &\times\left[
    e_{1}\delta(\textbf{x}-\textbf{Y}
    -\dfrac{m_{2}}{M}\textbf{y})
    +e_{2}\delta(\textbf{x}-\textbf{Y}
    -\dfrac{m_{1}}{M}\textbf{y})
    \right]
    \\
    &\times
    \left(
    \hat{\varphi}^{+}(\textbf{y},\textbf{Y})
    \dfrac{\partial\hat{\varphi}
    (\textbf{y},\textbf{Y})}{\partial\textbf{Y}}
    -\dfrac{\partial\hat{\varphi}^{+}
    (\textbf{y},\textbf{Y})}{\partial\textbf{Y}}
    \hat{\varphi}(\textbf{y},\textbf{Y})
    \right).
\end{split}
\end{equation}
As is easy to see, the operators $\hat{\sigma}_{0}(\textbf{x})$,
$\hat{\textbf{j}}_{0}(\textbf{x})$ in these expressions define the
bound states contribution to the charge and current densities.

In the momentum representation
\begin{equation*}
\begin{split}
    &\hat{\chi}_{i}(\textbf{x})
    ={1\over\sqrt{V}}\sum_\textbf{p}
    e^{i\textbf{p}\textbf{x}}
    \hat{a}_{i\textbf{p}},\quad i=1,2,
    \\
    &\hat{\eta}_{\alpha}(\textbf{x})
    ={1\over\sqrt{V}}\sum_\textbf{p}
    e^{i\textbf{p}\textbf{x}}
    \hat{\eta}_{\alpha}(\textbf{p}),
\end{split}
\end{equation*}
expressions \eqref{eq:16} according to \eqref{eq:08} will have the
next form:
\begin{equation}\label{eq:17}                              %%%% ---- %%%% Eq. 17
\begin{split}
    &\hat{\sigma}_{i}(\textbf{x})=
    \dfrac{e_{i}}{V}\sum\limits_{\textbf{p},\textbf{p}'}
    e^{i\textbf{x}(\textbf{p}'-\textbf{p})}
    \hat{a}_{i\textbf{p}}^{+}\hat{a}_{i\textbf{p}'},
    \\
    &\hat{\textbf{j}}_{i}(\textbf{x})=
    \dfrac{e_{i}}{2m_{i}V}\sum
    \limits_{\textbf{p},\textbf{p}'}
    e^{i\textbf{x}(\textbf{p}'-\textbf{p})}
    (\textbf{p}+\textbf{p}')
    \hat{a}_{i\textbf{p}}^{+}\hat{a}_{i\textbf{p}'},
    \\
    &\hat{\sigma}_{0}(\textbf{x})=\dfrac{1}{V}
    \sum\limits_{\textbf{p},\textbf{p}'}
    \sum\limits_{\alpha,\beta}
    e^{i\textbf{x}(\textbf{p}'-\textbf{p})}
    \\
    &\qquad\qquad\times
    \sigma_{\alpha\beta}(\textbf{p}-\textbf{p}')
    \hat{\eta}_{\alpha}^{+}(\textbf{p})
    \hat{\eta}_{\beta}(\textbf{p}'),
    \\
    &\hat{\textbf{j}}_{0}(\textbf{x})=\dfrac{1}{V}
    \sum\limits_{\textbf{p},\textbf{p}'}
    \sum\limits_{\alpha,\beta}
    e^{i\textbf{x}(\textbf{p}'-\textbf{p})}
    \left(\textbf{I}_{\alpha\beta}(\textbf{p}-\textbf{p}')
    \right.
    \\
    &\qquad\left.
    +\dfrac{(\textbf{p}+\textbf{p}')}{2M}
    \sigma_{\alpha\beta}(\textbf{p}-\textbf{p}')
    \right)
    \hat{\eta}_{\alpha}^{+}(\textbf{p})
    \hat{\eta}_{\beta}(\textbf{p}'),
\end{split}
\end{equation}
where (see \eqref{eq:08})
\begin{equation}\label{eq:18}                              %%%% ---- %%%% Eq. 18
\begin{split}
    &\sigma_{\alpha\beta}(\textbf{k})
    =\int d\textbf{y}\varphi_{\alpha}^{*}(\textbf{y})
    \varphi_{\beta}(\textbf{y})
    \\
    &\qquad\times\left[e_{1}\exp{i\dfrac{m_{2}}{M}
    \textbf{k}\textbf{y}}
    +e_{2}\exp{(-i\dfrac{m_{1}}{M}
    \textbf{k}\textbf{y})}\right],
    \\
    &\textbf{I}_{\alpha\beta}(\textbf{k})
    =-\dfrac{i}{2}\int d\textbf{y}
    \\
    &\qquad\times\left(\varphi_{\alpha}^{*}(\textbf{y})
    \dfrac{\partial\varphi_{\beta}(\textbf{y})}
    {\partial\textbf{y})}-
    \dfrac{\partial\varphi_{\alpha}^{*}(\textbf{y})}
    {\partial\textbf{y})}\varphi_{\beta}(\textbf{y})
    \right)
    \\
    &\quad\times
    \left[\dfrac{e_{1}}{m_{1}}\exp{i\dfrac{m_{2}}{M}
    \textbf{k}\textbf{y}}
    -\dfrac{e_{2}}{m_{2}}\exp{(-i\dfrac{m_{1}}{M}
    \textbf{k}\textbf{y})}\right].
\end{split}
\end{equation}

It is significant to note, that the Hamiltonian for free particles
$\hat{\mathcal H}_{p}$ (see \eqref{eq:01}, \eqref{eq:03}) in the
momentum representation can be written as
\begin{equation}\label{eq:19}                              %%%% ---- %%%% Eq. 19
\begin{split}
    &\qquad
    \hat{\mathcal H}_{p}=\hat{\mathcal H}_{1p}
    +\hat{\mathcal H}_{2p}+\hat{\mathcal H}_{0p},
    \\
    &\hat{\mathcal H}_{ip}=\sum\limits_{\textbf{p}}
    \varepsilon_{i}(\textbf{p})
    \hat{a}_{i\textbf{p}}^{+}\hat{a}_{i\textbf{p}},
    ~~~i=1,2,
    \\
    &\hat{\mathcal H}_{0p}=\sum\limits_{\alpha}
    \sum\limits_{\textbf{p}}
    \varepsilon_{\alpha}(\textbf{p})
    \hat{\eta}_{\alpha}(\textbf{p})^{+}
    \hat{\eta}_{\alpha}(\textbf{p}),
    \\
    &\varepsilon_{i}(\textbf{p})
    ={\textbf{p}^2}/{2m_i},
    \quad
    \varepsilon_{\alpha}(\textbf{p})
    =\varepsilon_{\alpha}+{\textbf{p}^2}/{2M},
    \end{split}
\end{equation}
where $\varepsilon_{\alpha}$ is the energy of the atomic level
with quantum numbers $\alpha$, $M$ is the bound state mass,
$M=(m_{1}+m_{2})$.

The Maxwell equations for our system according to \cite{ref:02}
can be written in the following form
\begin{equation}\label{eq:20}                              %%%% ---- %%%% Eq. 20
\begin{split}
    \dfrac{\partial\hat{\textbf{H}}}{\partial t}
    =-c\text{rot}\hat{\textbf{E}},\qquad
    \text{div}\hat{\textbf{E}}
    =4\pi(\hat{\sigma}+\sigma^{(e)}),
    \\
    \dfrac{\partial\hat{\textbf{E}}}{\partial t}
    =c\text{rot}\hat{\textbf{H}}
    -4\pi(\hat{\textbf{J}}+\textbf{J}^{(e)}),\quad
    \text{div}\hat{\textbf{H}}=0,
\end{split}
\end{equation}
where operators $\hat\sigma$, $\hat{\textbf{J}}$ are still defined
by the expressions \eqref{eq:10}, \eqref{eq:12}, \eqref{eq:13} and
values $\sigma^{(e)}$, ${\textbf{J}}^{(e)}$ are the external current
and charge densities. The electric $\hat{\textbf{E}}$ and magnetic
$\hat{\textbf{H}}$ field intensity operators in terms of the scalar
and vector potentials can be expressed as (see \cite{ref:01},
\cite{ref:02} and also equations \eqref{eq:10}, \eqref{eq:11})
\begin{equation}\label{eq:21}                              %%%% ---- %%%% Eq. 21
\begin{split}
    &\hat{\textbf{H}}=rot{\hat{\textbf{A}}},
    \\
    &\hat{\textbf{E}}
    =-{1\over c}{\partial{\hat{\textbf{A}}}
    \over\partial t} -{\partial\over\partial\textbf{x}}
    \left(\varphi^{(e)}+\int d\textbf{x}'{{\hat\sigma}
    (\textbf{x}')\over\vert\textbf{x}
    -\textbf{x}'\vert} \right).
\end{split}
\end{equation}
Note that in deriving the electrodynamic equations we used the
Coulomb's gauge.
\section{The system response to the perturbation by the external
electromagnetic field}

In this section to study the system response to the perturbing
action of the external electromagnetic field we will follow the
principles that have been stated in \cite{ref:01}. Let us consider a
system that at some moment of time $t$ is characterized by
statistical operator $\rho(t)$. Considering that the Hamiltonian of
interaction $\hat{V}(t)$ is linear in respect to the external field
and assuming that it is small in comparison with the Hamiltonian
$\hat{\mathcal H}=\hat{\mathcal H}_{0}+\hat{\mathcal H}_{int}$ (see
\eqref{eq:01}), we can develop the perturbation theory over the week
interaction. In accordance with \cite{ref:01} the mean value of an
arbitrary quasilocal operator ${\hat a}(\textbf{x})$ in linear
approach for such system can be written as
\begin{equation}\label{eq:22}                              %%%% ---- %%%% Eq. 22
\begin{split}
    &\text{Sp}\rho(t){\hat a}(\textbf{x})
    =\text{Sp}w{\hat a}(0)+a^{F}(\textbf{x},t),
    \\
    &a^{F}(\textbf{x},t)=\int_{-\infty}^{\infty}
    dt'\int d\textbf{x}'
    \\
    &\qquad\quad\times G_{a\xi_{i}}^{(+)}
    ({\textbf{x}}-\textbf{x}',t-t')
    F_{i}(\textbf{x}',t'),
\end{split}
\end{equation}
where $w$ is the Hibbs distribution operator
\begin{equation}\label{eq:23}                              %%%% ---- %%%% Eq. 23
    w=\exp\{\Omega-\beta
    (\hat{\mathcal H}-\mu_{1}{\hat N}_{1}
    -\mu_{2}{\hat N}_{2})\},
\end{equation}
$\beta={1/T}$ is the reciprocal temperature, ${\hat N}_{1}$, ${\hat N}_{2}$ are
the density operators of all fermions of the first and second kind (including
fermions in bound states, see \eqref{eq:12})
\begin{equation}\label{eq:24}                              %%%% ---- %%%% Eq. 24
    {\hat N}_{1}=\int{d\textbf{x}}
    \hat{\rho}_{1}(\textbf{x}),\quad
    {\hat N}_{2}=\int{d\textbf{x}}
    \hat{\rho}_{2}(\textbf{x})
\end{equation}
and $\mu_{1}$, $\mu_{2}$ are the chemical potentials of fermions of
the first and second kind. The thermodynamic parameters $\beta$,
$\mu_{1}$, $\mu_{2}$ can be found from the relations
\begin{equation}\label{eq:25}                              %%%% ---- %%%% Eq. 25
    \text{Sp}w\hat{\mathcal H}
    ={\mathcal H},\quad
    \text{Sp}w{\hat N}_{1}=N_{1},
    \quad \text{Sp}w{\hat N}_{2}=N_{2},
\end{equation}
and the thermodynamic potential $\Omega$ dependence on thermodynamic
parameters is defined by the expression
\begin{equation*}                                         %%%% ---- %%%%
    \text{Sp}w=1.
\end{equation*}

In the formula \eqref{eq:22} $F_{i}(\textbf{x},t)$ are the
quantities, that define the external field and
${\hat\xi}_{i}(\textbf{x})$ are quasilocal operators, related to our
system (see also \cite{ref:01}); the summation convention is assumed
for the repeated index $i$.

And, finally, the quantity $G_{a\xi_{i}}^{(+)}
({\textbf{x}}-\textbf{x}',t-t')$ in the expression \eqref{eq:22} is
the two-time retarded Green function (note, that ''tilde'' over
operators means that they are taken in the Heisenberg
representation)

\begin{equation}\label{eq:26}                              %%%% ---- %%%% Eq. 26
\begin{split}
    G_{a\xi_{i}}^{(+)}({\textbf{x}}-\textbf{x}',t-t')
    =-i\theta(t-t')\quad
    \\
    \times Spw[{\tilde{\hat
    a}}(\textbf{x},t),
    \tilde{\hat{\xi_{i}}}(\textbf{x}',t')],
\end{split}
\end{equation}
where $\theta(t)$ is Heaviside function
\begin{equation*}                                          %%%% ---- %%%%
    \theta(t)=
    \begin{cases}
    1,\  t>0,
    \\
    0,\  t<0.
    \end{cases}
\end{equation*}

Going over to Fourier transforms of values $a^{F}$, $F_{i}$
\begin{equation*}                                         %%%% ---- %%%%
\begin{split}
    a^{F}(\textbf{x},t)
    ={1\over(2\pi)^{4}}
    \int d{\textbf{k}}d\omega
    e^{-i(t\omega-{\textbf{k}}
    {\textbf{x}})}a^{F}(\textbf{k},\omega),
    \\
    F_{i}(\textbf{x},t)={1\over(2\pi)^{4}}
    \int d{\textbf{k}}d\omega e^{-i(t\omega
    -{\textbf{k}}{\textbf{x}})}
    F_{i}(\textbf{k},\omega)
\end{split}
\end{equation*}
one obtains
\begin{equation}\label{eq:27}                              %%%% ---- %%%% Eq. 27
    a^{F}(\textbf{k},\omega)
    =G_{a\xi_{i}}^{(+)}({\textbf{k}},\omega)
    F_{i}(\textbf{k},\omega),
\end{equation}
where
\begin{equation}\label{eq:28}                              %%%% ---- %%%% Eq. 28
    G_{a\xi_{i}}^{(+)}({\textbf{k}},\omega)
    =\int_{-\infty}^{\infty}dt\int
    d\textbf{x}e^{i(t\omega-{\textbf{k}}
    {\textbf{x}})}G_{a\xi_{i}}^{(+)}({\textbf{x}},t).
\end{equation}

It is significant, that in terms of the Fourier transforms of the
introduced quantities we can express also the energy, transferred
from field to matter. If to assume, that field is acting only for a
limited period of time, the total energy $Q$, received by matter, is
given by the expression \cite{ref:01}
\begin{equation}\label{eq:29}                              %%%% ---- %%%% Eq. 29
\begin{split}
    Q={i\over (2\pi)^{4}}
    \int_{-\infty}^{\infty}d\omega
    \int d\textbf{k}\omega\qquad\qquad
    \\
    \times F_{i}(-\textbf{k},-\omega)G_{\xi_{i}
    \xi_{j}}^{(+)}({\textbf{k}},\omega)
    F_{j}(\textbf{k},\omega).
\end{split}
\end{equation}

Now we can apply these expressions for studying the response of the
system, consisting of two kinds of oppositely charged fermions and
their bound states. To make use of the Green function method, that
was described above (see \eqref{eq:22}--\eqref{eq:28}), it is more
convenient to represent the system Hamiltonian, that is defined by
formulas \eqref{eq:01}--\eqref{eq:16}, as
\begin{equation}\label{eq:30}                              %%%% ---- %%%% Eq. 30
    \hat{\mathcal H}(t)
    =\hat{\mathcal H}+{\hat V}^{(e)}(t),\quad
    \hat{\mathcal H}=\hat{\mathcal H}_{0}+
    \hat{\mathcal H}_{int}+{\hat V},
\end{equation}
where $\hat{\mathcal H}_{0}$ and $\hat{\mathcal H}_{int}$ are given
by the formulas \eqref{eq:01}--\eqref{eq:07}, ${\hat V}$ is defined
by the expression (see also \eqref{eq:10}--\eqref{eq:14})
\begin{equation}\label{eq:31}                              %%%% ---- %%%% Eq. 31
\begin{split}
    \hat{V}
    =-&\dfrac{1}{c}\int d\textbf{x}\hat{\textbf{a}}(\textbf{x},t)
    \hat{\textbf{j}}(\textbf{x},t)
    \\
    &-\dfrac{1}{2c^{2}}
    \int d\textbf{x}\hat{\textbf{a}}^{2}(\textbf{x},t)
    \sum\limits_{i=1}^{2}\dfrac{e_{i}}{m_{i}}\hat{\sigma}_{i}
    (\textbf{x}),
\end{split}
\end{equation}
and the Hamiltonian ${\hat V}^{(e)}(t)$ describes the system
interaction with the external electromagnetic field
\begin{equation}\label{eq:32}                              %%%% ---- %%%% Eq. 32
\begin{split}
    {\hat V}^{(e)}(t)&=-\dfrac{1}{c}\int
    d\textbf{x}\textbf{A}^{(e)}(\textbf{x},t)
    \hat{\textbf{j}}(\textbf{x})
    \\
    +&\dfrac{1}{2c^{2}}
    \int d\textbf{x}\textbf{A}^{(e)}(\textbf{x},t)^{2}
    \sum\limits_{i=1}^{2}\dfrac{e_{i}}{m_{i}}\hat{\sigma}_{i}
    (\textbf{x})
    \\
    &+\int d\textbf{x}\varphi^{(e)}(\textbf{x},t)
    \hat{\sigma}(\textbf{x}).
\end{split}
\end{equation}

To get the Maxwell equations for the electromagnetic field in
medium, it is necessary to average the equations \eqref{eq:20} with
the system statistical operator containing the information both
about medium and electromagnetic field. To this end we will define
the mean values of the electromagnetic fields
$\textbf{E}(\textbf{x},t)$, $\textbf{H}(\textbf{x},t)$, acting in
the matter
\begin{equation}\label{eq:33}                              %%%% ---- %%%% Eq. 33
    \textbf{E}(\textbf{x},t)
    =\text{Sp}\rho(t){\hat{\textbf{E}}}(\textbf{x},t),~
    \textbf{H}(\textbf{x},t)
    =\text{Sp}\rho(t){\hat{\textbf{H}}}(\textbf{x},t),
\end{equation}
and also the induced charge and current averages (see
\eqref{eq:12}--\eqref{eq:14})
\begin{equation}\label{eq:34}                              %%%% ---- %%%% Eq. 34
    \textbf{J}(\textbf{x},t)
    =\text{Sp}\rho(t)\hat{\textbf{J}}(\textbf{x},t),~
    \sigma(\textbf{x},t)
    =\text{Sp}\rho(t){\hat\sigma}(\textbf{x}).
\end{equation}
The equations \eqref{eq:20}, averaged in accordance with the
formulas \eqref{eq:33}--\eqref{eq:34}, bring us to the
Maxwell--Lorentz equations for the average fields in the matter
\begin{equation}\label{eq:35}                              %%%% ---- %%%% Eq. 35
\begin{split}
    \dfrac{\partial\textbf{H}}{\partial t}
    =-c\text{rot}\textbf{E},\qquad
    \text{div}\textbf{E}
    =4\pi(\sigma+\sigma^{(e)}),
    \\
    \dfrac{\partial\textbf{E}}{\partial t}
    =c\text{rot}\textbf{H}
    -4\pi(\textbf{J}+\textbf{J}^{(e)}),~~
    \text{div}\textbf{H}=0.
\end{split}
\end{equation}
where quantities $\sigma^{(e)}$, ${\textbf{J}}^{(e)}$ still
represent the extrinsic charge and current densities.

The next problem is to find the charge $\sigma(\textbf{x},t)$ and
current $\textbf{J}(\textbf{x},t)$ densities induced by the external
field. Calculating these quantities under assumption of week
interaction between the system and the external field we will use
the equations \eqref{eq:22}--\eqref{eq:28}, considering the
potentials $\textbf{A}^{(e)}(\textbf{x},t)$,
$\varphi^{(e)}(\textbf{x},t)$ as $F_{i}(\textbf{x},t)$, and
${\hat\sigma}(\textbf{x})$ or $\hat{\textbf{J}}(\textbf{x},t)$ as a
quasilocal operator ${\hat a}(\textbf{x})$. As a result one gets
\begin{equation}\label{eq:36}                              %%%% ---- %%%% Eq. 36
\begin{split}
    &\widetilde{\sigma}(\textbf{x},t)
    =\sum\limits_{a}\sigma_{a}
    +\int\limits_{-\infty}^{\infty}dt'
    \int d^{3}x'
    \\
    &\qquad\qquad\times\left[
    -\overline{G}^{(+)}_{i}(\textbf{x}-\textbf{x}',t-t')
    \dfrac{1}{c}A_{i}^{(e)}(\textbf{x}',t')\right.
    \\
    &\qquad\qquad~~\left.
    +G^{(+)}(\textbf{x}-\textbf{x}',t-t')
    \varphi^{(e)}(\textbf{x}',t')
    \right],
    \\
    &\widetilde{J}_{k}(\textbf{x},t)
    =-\dfrac{1}{c}A_{k}^{(e)}(\textbf{x},t)
    \sum\limits_{a}\dfrac{e_{a}}{m_{a}}\sigma_{a}
    \\
    &+\int\limits_{-\infty}^{\infty}dt'
    \int d^{3}x'\left[
    -G^{(+)}_{kl}(\textbf{x}-\textbf{x}',t-t')
    \dfrac{1}{c}A_{l}^{(e)}(\textbf{x}',t')
    \right.
    \\
    &\qquad\left.
    +G^{(+)}_{k}(\textbf{x}-\textbf{x}',t-t')
    \varphi^{(e)}(\textbf{x}',t')
    \right],
\end{split}
\end{equation}
where $\sigma_{a}=\text{Sp}w\hat{\sigma}_{a}(0)$, $a=1,2,0$ (see
\eqref{eq:17}), $w$ is the Hibbs statistical operator~\eqref{eq:23}.
It is significant to note, that for quasineutral systems, where the
number of fermions are equal ($N_{1}=N_{2}$, see \eqref{eq:24}) and
their absolute charge values are also equal ($\vert e_{1}\vert=\vert
e_{2}\vert$) $\sum\limits_{a}\sigma_{a}=0$.

The retarded charge and current Green functions, that are included
in the expression \eqref{eq:36}, are determined in accordance with
the formula \eqref{eq:26} (see also \cite{ref:01}):
\begin{equation}\label{eq:37}                              %%%% ---- %%%% Eq. 37
\begin{split}
    G^{(+)}(\textbf{x},t)=-i\theta(t)\text{Sp}
    {w[\hat{\sigma}(\textbf{x},t),\hat{\sigma}(0)]},
    \\
    G^{(+)}_{k}(\textbf{x},t)=-i\theta(t)\text{Sp}
    {w[\hat{j}_{k}(\textbf{x},t),\hat{\sigma}(0)]},
    \\
    \overline{G}^{(+)}_{k}(\textbf{x},t)=-i\theta(t)\text{Sp}
    {w[\hat{\sigma}(\textbf{x},t),\hat{j}_{k}(0)]},
    \\
    G^{(+)}_{kl}(\textbf{x},t)=-i\theta(t)\text{Sp}
    {w[\hat{j}_{k}(\textbf{x},t),\hat{j}_{l}(0)]}.
\end{split}
\end{equation}
As the charge and current density operators of particles of
different kinds (see \eqref{eq:17}) commute with each other
\begin{equation*}                                             %%%% ---- %%%%
    [\hat{\sigma}_{a},\hat{\sigma}_{b}]=
    [\hat{\sigma}_{a},\hat{\textbf{j}}_{b}]=
    [\hat{\textbf{j}}_{a},\hat{\textbf{j}}_{b}]=0,~
    a\neq b,~a,b=1,2,0,
\end{equation*}
then, according to the equations \eqref{eq:15}, \eqref{eq:17},
\eqref{eq:37}, the contribution of different kinds of particles to
Green functions will be additive
\begin{equation}\label{eq:38}                              %%%% ---- %%%% Eq. 38
\begin{split}
    &G^{(+)}(\textbf{x},t)=\sum_{a}G^{(+)}_{a}(\textbf{x},t),
    \\
    &\qquad G^{(+)}_{a}(\textbf{x},t)=
    -i\theta(t)\text{Sp}{w[\hat{\sigma}_{a}
    (\textbf{x},t),\hat{\sigma}_{a}(0)]},
    \\
    &\overline{G}^{(+)}_{k}(\textbf{x},t)=\sum_{a}
    \overline{G}^{(+)}_{ak}(\textbf{x},t),
    \\
    &\qquad \overline{G}^{(+)}_{ak}(\textbf{x},t)=
    -i\theta(t)\text{Sp}{w[
    \hat{\sigma}_{a}(\textbf{x},t),\hat{j}_{ak}(0)]},
    \\
    &G^{(+)}_{k}(\textbf{x},t)
    =\sum_{a}G^{(+)}_{ak}(\textbf{x},t),
    \\
    &\qquad G^{(+)}_{ak}(\textbf{x},t)=
    -i\theta(t)\text{Sp}{w[
    \hat{j}_{ak}(\textbf{x},t),\hat{\sigma}_{a}(0)]},
    \\
    &G^{(+)}_{kl}(\textbf{x},t)
    =\sum_{a}G^{(+)}_{akl}(\textbf{x},t),
    \\
    &\qquad G^{(+)}_{akl}(\textbf{x},t)=
    -i\theta(t)\text{Sp}{w[
    \hat{j}_{ak}(\textbf{x},t),\hat{j}_{al}(0)]}.
\end{split}
\end{equation}
With the help of direct calculations, following the method
\cite{ref:01}, we can see, that also in the presence of particle
bound states the following correspondence between Green functions
\eqref{eq:37} takes place
\begin{equation}\label{eq:39}                              %%%% ---- %%%% Eq. 39
\begin{split}
    &\overline{G}^{(+)}_{k}(\textbf{x},t)
    ={G}^{(+)}_{k}(\textbf{x},t),
    \\
    &\dfrac{\partial G_{i}^{(+)}
    (\textbf{x},t)}{\partial x_{i}}
    +\dfrac{\partial G^{(+)}
    (\textbf{x},t)}{\partial t}=0,
    \\
    &\dfrac{\partial G_{ki}^{(+)}
    (\textbf{x},t)}{\partial x_{k}}
    +\dfrac{\partial G_{i}^{(+)}
    (\textbf{x},t)}{\partial t}
    \\
    &\qquad+\sum\limits_{a}\dfrac{e_{a}}{m_{a}}
    \sigma_{a}\delta(t)
    \dfrac{\partial}{\partial x_{i}}
    \delta(\textbf{x})=0.
\end{split}
\end{equation}
For the Green functions Fourier transforms (see \eqref{eq:28}) these
relations can be written as
\begin{equation}\label{eq:40}                              %%%% ---- %%%% Eq. 40
\begin{split}
    &\overline{G}^{(+)}_{k}(\textbf{k},\omega)
    ={G}^{(+)}_{k}(\textbf{k},\omega),
    \\
    &G_{i}^{(+)}
    (\textbf{k},\omega)k_{i}
    -\omega G^{(+)}(\textbf{k},\omega)=0,
    \\
    G_{ij}^{(+)}
    (\textbf{k},\omega)&k_{j}
    -\omega G_{i}^{(+)}(\textbf{k},\omega)
    +k_{i}\sum\limits_{a}
    \dfrac{e_{a}}{m_{a}}\sigma_{a}=0.
\end{split}
\end{equation}
\section{Green functions and macroscopic characteristics of the
ideal low-temperature hydrogen-like plasma}

If we neglect of all interactions between particles in the
investigated system, it can be considered as an ideal hydrogen-like
low-temperature plasma (we note, that kinetic energy of particles
must be small in comparison with the binding energy of compound
particles). For an ideal hydrogen-like plasma the Green functions,
that was introduced earlier, can be calculated exactly. To do it we
must take into consideration, that neglecting the quantum fields
presence, the Hamiltonian $\hat{\mathcal H}$ in the formula
\eqref{eq:30} must be interpreted as $\hat{\mathcal H}_{p}$, see
\eqref{eq:01}, \eqref{eq:03}, \eqref{eq:19}. Taking into account
this fact the Heisenberg representation of charge and current
density operators, that appear in the equations \eqref{eq:37} for
the Green functions, is defined by expressions:
\begin{equation}\label{eq:41}                              %%%% ---- %%%% Eq. 41
\begin{split}
    \hat{\sigma}_{i}(\textbf{x},t)
    &=\dfrac{e_{i}}{V}\sum\limits_{\textbf{p},\textbf{p}'}
    e^{-i\textbf{x}(\textbf{p}-\textbf{p}')}
    e^{-it(\varepsilon_{i}(\textbf{p})
    -\varepsilon_{i}(\textbf{p}'))}
    \hat{a}_{i\textbf{p}}^{+}\hat{a}_{i\textbf{p}'},
    \\
    \hat{\textbf{j}}_{i}(\textbf{x},t)
    &=\dfrac{1}{2m_{i}}\hat{\sigma}_{i}(\textbf{x},t),
    \qquad i=1,2,
    \\
    \hat{\sigma}_{0}(\textbf{x},t)
    &=\dfrac{1}{V}\sum\limits_{\textbf{p},\textbf{p}'}
    \sum\limits_{\alpha,\beta}
    e^{-i\textbf{x}(\textbf{p}-\textbf{p}')}
    e^{-it(\varepsilon_{\alpha}(\textbf{p})
    -\varepsilon_{\beta}(\textbf{p}'))}
    \\
    &\times\sigma_{\alpha\beta}
    (\textbf{p}-\textbf{p}')
    \hat{\eta}_{\alpha}^{+}(\textbf{p})
    \hat{\eta}_{\beta}(\textbf{p}'),
    \\
    \hat{\textbf{j}}_{0}(\textbf{x},t)
    &=\dfrac{1}{V}\sum\limits_{\textbf{p},\textbf{p}'}
    \sum\limits_{\alpha,\beta}
    e^{-i\textbf{x}(\textbf{p}-\textbf{p}')}
    e^{-it(\varepsilon_{\alpha}(\textbf{p})
    -\varepsilon_{\beta}(\textbf{p}'))}
    \\
    &\times
    \left[\dfrac{(\textbf{p}+\textbf{p}')}{2M}
    \sigma_{\alpha\beta}(\textbf{p}-\textbf{p}')
    \right.
    \\
    &\qquad\left.
    +\textbf{I}_{\alpha\beta}(\textbf{p}-\textbf{p}')
    \right]
    \hat{\eta}_{\alpha}^{+}(\textbf{p})
    \hat{\eta}_{\beta}(\textbf{p}'),
\end{split}
\end{equation}
where quantities $\sigma_{\alpha\beta}(\textbf{k})$,
$\textbf{I}_{\alpha\beta}(\textbf{k})$ are given by the formulas
\eqref{eq:18}. If to substitute the operators \eqref{eq:41} in
\eqref{eq:38} and to do some calculations, we will come to the
following expressions for the Fourier transforms of scalar Green
functions (see \eqref{eq:28}):
\begin{equation}\label{eq:42}                              %%%% ---- %%%% Eq. 42
\begin{split}
    G^{(+)}_{1}(\textbf{k},\omega)
    &=\dfrac{e_{1}^{2}}{V}
    \sum\limits_{\textbf{p}}
    \dfrac{f_{1}(\textbf{p}-\textbf{k})-f_{1}(\textbf{p})}
    {\varepsilon_{1}(\textbf{p})-
    \varepsilon_{1}(\textbf{p}-\textbf{k})
    +\omega+i0},
    \\
    G^{(+)}_{2}(\textbf{k},\omega)
    &=\dfrac{e_{2}^{2}}{V}
    \sum\limits_{\textbf{p}}
    \dfrac{f_{2}(\textbf{p}-\textbf{k})-f_{2}(\textbf{p})}
    {\varepsilon_{2}(\textbf{p})-
    \varepsilon_{2}(\textbf{p}-\textbf{k})
    +\omega+i0},
    \\
    G^{(+)}_{0}(\textbf{k},\omega)
    &=\dfrac{1}{V}
    \sum\limits_{\textbf{p}}
    \sum\limits_{\alpha,\beta}
    \sigma_{\alpha\beta}(\textbf{k})
    \sigma_{\beta\alpha}(-\textbf{k})
    \\
    &\qquad\times
    \dfrac{f_{\alpha}(\textbf{p}-\textbf{k})
    -f_{\beta}(\textbf{p})}
    {\varepsilon_{\alpha}(\textbf{p})-
    \varepsilon_{\beta}(\textbf{p}-\textbf{k})
    +\omega+i0}.
\end{split}
\end{equation}
Similar expressions for the vector Green functions have the form:
\begin{equation}\label{eq:43}                              %%%% ---- %%%% Eq. 43
\begin{split}
    G^{(+)}_{1l}(\textbf{k},\omega)
    &=\dfrac{e_{1}^{2}}{2m_{1}V}
    \sum\limits_{\textbf{p}}
    (2\textbf{p}-\textbf{k})_{l}
    \\
    &\times
    \dfrac{f_{1}(\textbf{p}-\textbf{k})-f_{1}(\textbf{p})}
    {\varepsilon_{1}(\textbf{p})-
    \varepsilon_{1}(\textbf{p}-\textbf{k})
    +\omega+i0},
    \\
    G^{(+)}_{2l}(\textbf{k},\omega)
    &=\dfrac{e_{2}^{2}}{2m_{2}V}
    \sum\limits_{\textbf{p}}
    (2\textbf{p}-\textbf{k})_{l}
    \\
    &\times
    \dfrac{f_{2}(\textbf{p}-\textbf{k})-f_{2}(\textbf{p})}
    {\varepsilon_{2}(\textbf{p})-
    \varepsilon_{2}(\textbf{p}-\textbf{k})
    +\omega+i0},
    \\
    G^{(+)}_{0l}(\textbf{k},\omega)
    &=\dfrac{1}{V}
    \sum\limits_{\textbf{p}}
    \sum\limits_{\alpha,\beta}
    \left[\dfrac{(2\textbf{p}-\textbf{k})}{2M}
    \sigma_{\alpha\beta}(\textbf{k})
    +\textbf{I}_{\alpha\beta}(\textbf{k})
    \right]_{l}
    \\
    &\times
    \dfrac{\sigma_{\beta\alpha}(-\textbf{k})
    \left[f_{\alpha}(\textbf{p}-\textbf{k})
    -f_{\beta}(\textbf{p})\right]}
    {\varepsilon_{\alpha}(\textbf{p})-
    \varepsilon_{\beta}(\textbf{p}-\textbf{k})
    +\omega+i0}.
\end{split}
\end{equation}
And, finally, the tensor Green functions for the investigated
system is given by expressions:
\begin{equation}\label{eq:44}                              %%%% ---- %%%% Eq. 44
\begin{split}
    &G^{(+)}_{1ls}(\textbf{k},\omega)
    =\dfrac{e_{1}^{2}}{4m_{1}^{2}V}
    \sum\limits_{\textbf{p}}
    (2\textbf{p}-\textbf{k})_{l}
    (2\textbf{p}-\textbf{k})_{s}
    \\
    &\qquad\times\dfrac{f_{1}(\textbf{p}
    -\textbf{k})-f_{1}(\textbf{p})}
    {\varepsilon_{1}(\textbf{p})-
    \varepsilon_{1}(\textbf{p}-\textbf{k})
    +\omega+i0},
    \\
    &G^{(+)}_{2ls}(\textbf{k},\omega)
    =\dfrac{e_{2}^{2}}{4m_{2}^{2}V}
    \sum\limits_{\textbf{p}}
    (2\textbf{p}-\textbf{k})_{l}
    (2\textbf{p}-\textbf{k})_{s}
    \\
    &\qquad\times\dfrac{f_{2}(\textbf{p}
    -\textbf{k})-f_{2}(\textbf{p})}
    {\varepsilon_{2}(\textbf{p})-
    \varepsilon_{2}(\textbf{p}-\textbf{k})
    +\omega+i0},
    \\
    &G^{(+)}_{0lj}(\textbf{k},\omega)
    =\dfrac{1}{V}
    \sum\limits_{\textbf{p}}
    \sum\limits_{\alpha,\beta}
    \left[
    \dfrac{(2\textbf{p}-\textbf{k})}{2M}
    \sigma_{\alpha\beta}(\textbf{k})
    +\textbf{I}_{\alpha\beta}(\textbf{k})
    \right]_{l}
    \\
    &\times
    \left[
    \dfrac{(2\textbf{p}-\textbf{k})}{2M}
    \sigma_{\beta\alpha}(-\textbf{k})
    +\textbf{I}_{\beta\alpha}(-\textbf{k})
    \right]_{j}
    \\
    &\qquad\times
    \dfrac{f_{\alpha}(\textbf{p}-\textbf{k})
    -f_{\beta}(\textbf{p})}
    {\varepsilon_{\alpha}(\textbf{p})-
    \varepsilon_{\beta}(\textbf{p}-\textbf{k})
    +\omega+i0}.
\end{split}
\end{equation}
In the expressions \eqref{eq:42}--\eqref{eq:44} we have introduced
the distribution functions for free fermions of the first
$f_{1}(\textbf{p})$ and second $f_{2}(\textbf{p})$ kind, and also
the distribution functions $f_{\alpha}(\textbf{p})$ for
hydrogen-like atoms (bound states) with the set of quantum numbers
$\alpha$
\begin{equation}\label{eq:45}                              %%%% ---- %%%% Eq. 45
\begin{split}
    f_{i}(\textbf{p})
    =\{\exp[(\varepsilon_{i}
    (\textbf{p})-\mu_{i})/T]+1
    \}^{-1},
    \\
    f_{\alpha}(\textbf{p})=
    \{\exp[(\varepsilon_{\alpha}(\textbf{p})
    -\mu_{\alpha})/T]-1\}^{-1}
\end{split}
\end{equation}
in accordance with the equations
\begin{equation}\label{eq:46}                              %%%% ---- %%%% Eq. 46
\begin{split}
    \text{Sp}w\hat{a}_{i\textbf{p}}^{+}
    \hat{a}_{i\textbf{p}'}
    =\delta_{\textbf{p},\textbf{p}'}
    f_{i}(\textbf{p}),~i=1,2,
    \\
    \text{Sp}w\hat{\eta}_{\alpha}^{+}(\textbf{p})
    \hat{\eta}_{\beta}(\textbf{p}')
    =\delta_{\alpha,\beta}
    \delta_{\textbf{p},\textbf{p}'}
    f_{\alpha}(\textbf{p}).
\end{split}
\end{equation}
The particle energies $\varepsilon_{1,2}(\textbf{p})$,
$\varepsilon_{\alpha}(\textbf{p})$ in formulas
\eqref{eq:42}--\eqref{eq:45} are given by the expressions
\eqref{eq:19}, and values $\delta_{\alpha,\beta}$,
$\delta_{\textbf{p},\textbf{p}'}$ in \eqref{eq:46} are the Kronecker
symbols.

The particular feature of the obtained Green functions is that the
contribution of bound states in the processes under consideration
now is taken into account.

The expressions for Green functions that have been found allow us to
get an expressions for the matter macroscopic parameters, such as
conductivity, permittivity and magnetic permeability. To this end we
will also use the method described in \cite{ref:01}.

In accordance with the formulas \eqref{eq:35}, \eqref{eq:36}the next
relation between Fourier transforms takes place
\begin{equation}\label{eq:47}                              %%%% ---- %%%% Eq. 47
\begin{split}
    \widetilde{J}_{i}(\textbf{k},\omega)
    =&\bar{\sigma}^{l}(\textbf{k},\omega)
    k_{i}\dfrac{\textbf{k}
    \textbf{E}^{(e)}(\textbf{k},\omega)}{k^2}
    \\
    &+\bar{\sigma}^{t}(\textbf{k},\omega)
    \dfrac{\left[[\textbf{k},
    \textbf{E}^{(e)}(\textbf{k},\omega)],
    \textbf{k}\right]}{k^2}~,
    \\
    \widetilde{\sigma}(\textbf{k},\omega)
    =&\dfrac{1}{\omega}
    \bar{\sigma}^{l}(\textbf{k},\omega)
    \textbf{k}\textbf{E}^{(e)}(\textbf{k},\omega),
\end{split}
\end{equation}
where
\begin{equation}\label{eq:48}                              %%%% ---- %%%% Eq. 48
\begin{split}
    \bar{\sigma}^{l}(\textbf{k},\omega)
    =&\dfrac{i\omega}{k^{2}}
    G^{(+)}(\textbf{k},\omega),
    \\
    \bar{\sigma}^{t}(\textbf{k},\omega)
    =&\dfrac{i}{\omega}\left[
    \sum\limits_{a}
    \dfrac{e_{a}}{m_{a}}\sigma_{a}\right.
    \\
    &\left.+\dfrac{1}{2}\left(\delta_{ij}
    -\dfrac{k_{i}k_{j}}{k^{2}}
    \right)G_{ij}
    ^{(+)}(\textbf{k},\omega)
    \right].
\end{split}
\end{equation}
It is clear from equations \eqref{eq:47} that the quantities
$\bar{\sigma}^{l}$ and $\bar{\sigma}^{t}$, expressed in terms of
Green functions according to formula \eqref{eq:48}, define the
longitudinal and transversal current density components. They are
usually interpreted as outer conductivity coefficients in contrast
to inner (or true) longitudinal $\sigma^l$ or transversal $\sigma^t$
conductivity coefficients, that will be defined below. Note that
according to \eqref{eq:38} these coefficients are also additive
quantities
\begin{equation*}                                          %%%% ---- %%%%
    \bar{\sigma}^{l,t}(\textbf{k},\omega)
    =\sum_{a}\bar{\sigma}^{l,t}_{a}(\textbf{k},\omega).
\end{equation*}

In terms of the introduced outer conductivity coefficients
$\bar{\sigma}^{l}$ and $\bar{\sigma}^{t}$ we can express the energy,
absorbed from the external field sources (see \eqref{eq:29}):
\begin{equation*}                                          %%%% ---- %%%%
    Q_{\omega \textbf{k}}
    =\dfrac{-2}{(2\pi)^4}\text{Im}{1\over \omega}
    \textbf{E}^{*(e)}(\textbf{k},\omega)
    G_{ij}^{+}(\textbf{k},\omega)
    \textbf{E}^{(e)}(\textbf{k},\omega).
\end{equation*}
From this expression, in accordance with the formulas
\eqref{eq:47}--\eqref{eq:48} one obtains
\begin{equation}\label{eq:49}                              %%%% ---- %%%% Eq. 49
\begin{split}
    Q_{\omega \textbf{k}}
    =\dfrac{2}{(2\pi)^4}\text{Re}\{
    \bar{\sigma}^{l}(\textbf{k},\omega)
    |\textbf{E}_{\parallel}^{(e)}(\textbf{k},\omega)|^2
    \\
    +\bar{\sigma}^{t}(\textbf{k},\omega)
    |\textbf{E}_{\perp}^{(e)}
    (\textbf{k},\omega)|^2\}.
\end{split}
\end{equation}

In terms of these quantities ($\bar{\sigma}^{l}$ and
$\bar{\sigma}^{t}$) the expressions for permittivity and magnetic
permeability can be also defined. The relation between the
permittivity and outer conductivity (see \cite{ref:01}) is given by
the formula:
\begin{equation*}                                          %%%% ---- %%%%
    \epsilon
    =\left(1+
    \dfrac{4\pi \bar{\sigma}^{l}}{i\omega}
    \right)^{-1}.
\end{equation*}
From this, according to the expression \eqref{eq:48}:
\begin{equation}\label{eq:50}                              %%%% ---- %%%% Eq. 50
    \epsilon^{-1}(\textbf{k},\omega)
    =1+\dfrac{4\pi}{k^2}
    G^{(+)}(\textbf{k},\omega).
\end{equation}
It is more convenient to express the magnetic permeability in terms
of inner conductivity coefficients $\sigma^l$ and $\sigma^t$
\begin{equation}\label{eq:51}                              %%%% ---- %%%% Eq. 51
    \mu^{-1}(\textbf{k},\omega)
    =1+
    \dfrac{4\pi\omega}{ic^2k^2}
    (\sigma^l-\sigma^t),
\end{equation}
that are connected with the outer conductivity coefficients
$\bar{\sigma}^{l}$ and $\bar{\sigma}^{t}$ by the relations:
\begin{equation*}                                          %%%% ---- %%%%
    \sigma^l=\varepsilon\bar{\sigma}^{l},
    ~~~~~~~\sigma^t=
    \dfrac{\bar{\sigma}^{t}}
    {1+\dfrac{4\pi \bar{\sigma}^{t}}{i\omega}
    \left(1-\dfrac{k^2c^2}{\omega^2}
    \right)^{-1}}.
\end{equation*}

In accordance with the expression \eqref{eq:49}, their relations
with the Green functions \eqref{eq:42}, \eqref{eq:44} take the form:
\begin{equation}\label{eq:52}                              %%%% ---- %%%% Eq. 52
\begin{split}
    &\sigma^l(\textbf{k},\omega)
    =\dfrac{i\omega G^{(+)}(\textbf{k},\omega)}
    {k^2+4\pi G^{(+)}(\textbf{k},\omega)},
    \\
    &\sigma^t(\textbf{k},\omega)
    =\dfrac{k^2 c^2-\omega^2}{i\omega}
    \dfrac{A(\textbf{k},\omega)}
    {(\omega^2-k^2 c^2)
    +4\pi A(\textbf{k},\omega)},
    \\
    &A(\textbf{k},\omega)
    \equiv\sum\limits_{a}
    \dfrac{e_{a}}{m_{a}}\sigma_{a}
    +\dfrac{1}{2}\left(\delta_{ij}
    -\dfrac{k_{i}k_{j}}{k^{2}}
    \right)G_{ij}
    ^{(+)}(\textbf{k},\omega).
\end{split}
\end{equation}

So, we have defined the main macroscopic characteristics of the
ideal hydrogen-like plasma in low temperature region. These
characteristics allow us to solve a number of applied problems for
our system. Let us demonstrate it on few examples.

Using the developed theory it is not difficult to find the
permittivity of an ideal gas of hydrogen-like (alkali) atoms at low
temperatures. According to the expressions \eqref{eq:42},
\eqref{eq:50} in neglect of free fermions contribution one gets
\begin{equation}\label{eq:53}                              %%%% ---- %%%% Eq. 53
\begin{split}
    \epsilon^{-1}(\textbf{k},\omega)
    =1+\dfrac{4\pi}{k^2}
    \dfrac{1}{V}
    \sum\limits_{\textbf{p}}
    \sum\limits_{\alpha,\beta}
    \sigma_{\alpha\beta}(\textbf{k})
    \sigma_{\beta\alpha}(-\textbf{k})
    \\
    \times
    \dfrac{f_{\alpha}(\textbf{p}-\textbf{k})
    -f_{\beta}(\textbf{p})}
    {\varepsilon_{\alpha}(\textbf{p})-
    \varepsilon_{\beta}(\textbf{p}-\textbf{k})
    +\omega+i0}.
\end{split}
\end{equation}

As good known, at extremely low temperatures the Bose-Einstein
condensate (BEC) of alkali atoms can be formed. At the temperatures
much lower the critical point temperature ($T\ll T_0$, see
\emph{e.g.} \cite{ref:01}) the bound states distribution functions
$f_\alpha (\textbf{p})$ are proportional to the Dirac delta-function
$\delta(\textbf{p})$. Therefore, according to the expressions
\eqref{eq:19}, \eqref{eq:53}, after integration over momentum
$\textbf{p}$ the expression for permittivity of the studied gas in
BEC state ($T\ll T_0$) take the form:
\begin{equation}\label{eq:54}                              %%%% ---- %%%% Eq. 54
\begin{split}
    \epsilon^{-1}(\textbf{k},\omega)
    \approx 1&+\dfrac{1}{2\pi^2 k^2}
    \sum\limits_{\alpha,\beta}
    \sigma_{\alpha\beta}(\textbf{k})
    \sigma_{\beta\alpha}(-\textbf{k})
    \\
    &\times\left[
    \dfrac{\nu_{\alpha}}
    {\omega+\Delta\varepsilon_{\alpha\beta}
    -\varepsilon_{k}+i\gamma_{\alpha\beta}}\right.
    \\
    &\qquad\left.-\dfrac{\nu_{\beta}}
    {\omega+\Delta\varepsilon_{\alpha\beta}
    +\varepsilon_{k}+i\gamma_{\alpha\beta}}\right],
\end{split}
\end{equation}
where $\nu_\alpha$ is the density of condensed atoms in the quantum
state $\alpha$, $\varepsilon_{k}=k^2/2M$ and quantities
$\sigma_{\alpha\beta}(\textbf{k})$ are still defined by the formula
\eqref{eq:18}. Note that due to the damping processes in real
systems we also introduced linewidth $\gamma_{\alpha\beta}$,
concerned with the transition probability from the state $\alpha$ to
the state $\beta$.

As it easy to see, in the expression \eqref{eq:54} at frequencies
that are close to the energy interval
$\Delta\varepsilon_{\alpha\beta}$
($\Delta\varepsilon_{\alpha\beta}\equiv\varepsilon_{\alpha}
-\varepsilon_{\beta}$) some peculiarities appear. In fact, such
behavior can strongly reflect on the dispersion characteristics of
the studied gas. It is a very interesting question, and we shall
return to it in the section~\ref{s5}.

Basing on the developed theory we can also find the energy, that is
dissipated by a charged particle when it passes through
hydrogen-like plasma at low temperature (see in that case, e.g.
\cite{ref:01}). In the case of a small dissipation the particle
movement can be considered as uniform. Thus, the particle current
density (the external current density in medium, see \eqref{eq:20})
will be defined by the formula
\begin{equation*}                                         %%%% ---- %%%%
    {\textbf{J}}^{(e)}({\textbf{x}},t)
    =ze{\textbf{v}}\delta({\textbf{x}}
    -{\textbf{v}}t),
\end{equation*}
where $ze$ is the particle charge and ${\textbf{v}}$ is the
particle velocity. It is easy to see, that the  Fourier transform
of the current density is given by the expression
\begin{equation}\label{eq:55}                              %%%% ---- %%%% Eq. 55
    {\textbf{J}}^{(e)}({\textbf{k}},\omega)
    =2\pi ze{\textbf{v}}\delta
    (\omega-{\textbf{k}}{\textbf{v}}).
\end{equation}
Next we will use the expression \eqref{eq:49} for the energy,
absorbed in the matter from external field sources. In this
expression, using \eqref{eq:47}, the Fourier transforms of the
longitudinal and transversal components of the external field can be
expressed in terms of longitudinal and transversal components of the
current particle density \eqref{eq:55}. If we do the necessary
calculations, we will come to the following expression for the
energy $d{\cal E}_{\textbf{k}\omega}$, that was dissipated by the
charged particle per unit time in the frequency $d\omega$ and the
wave vector $d\textbf{k}$ intervals, when it passes through the
hydrogen-like plasma:
\begin{equation*}                                         %%%% ---- %%%%
    d{\cal E}_{\textbf{k}\omega}
    =-q_{\textbf{k}\omega}d\omega d\textbf{k},
\end{equation*}
\begin{equation}\label{eq:56}                              %%%% ---- %%%% Eq. 56
\begin{split}
    q_{\textbf{k}\omega}
    ={Q_{\omega \textbf{k}}\over T}
    =-\left(ze\over 2\pi\right)^{2}\delta
    (\omega-{\textbf{k}}{\textbf{v}})\omega
    \\
    \times\text{Im}\left({{v^{2}\over
    c^{2}}-{1\over\epsilon\mu}}\right)
    \left({{\omega^{2}\over
    c^{2}}\epsilon-{k^{2}\over\mu}}
    \right)^{-1},
\end{split}
\end{equation}
where $T$ is the particle time of flight. To get the expression
\eqref{eq:56} it is necessary to use the formula
\begin{equation*}                                         %%%% ---- %%%%
    \delta^{2}(\omega-{\textbf{k}}{\textbf{v}})
    ={T\over2\pi}\delta(\omega
    -{\textbf{k}}{\textbf{v}}).
\end{equation*}
The total dissipated particle energy ${\cal E}$ per unit length can
be found by integrating the expression \eqref{eq:56} over $\omega$
and $\textbf{k}$
\begin{equation}\label{eq:57}                              %%%% ---- %%%% Eq. 57
    {d{\cal E}\over dx}=-{1\over v}\int d\omega
    d^{3}kq_{\textbf{k}\omega}.
\end{equation}
It is easy to see, that the main contribution in this integral comes
from poles of the integrand (see \eqref{eq:56})
\begin{equation}\label{eq:58}                              %%%% ---- %%%% Eq. 58
    \epsilon(\textbf{k},\omega)
    =0,\qquad {\omega^{2}\over
    c^{2}}\epsilon(\textbf{k},\omega)
    \mu(\textbf{k},\omega)-k^{2}=0.
\end{equation}
The formulas \eqref{eq:56}--\eqref{eq:58} are similar to the
expressions, that are given in \cite{ref:01}, however, they take an
account of particle bound states (atoms) to all processes that take
place in our system (see \eqref{eq:50}--\eqref{eq:52},
\eqref{eq:42}--\eqref{eq:46}). Note also that the expressions
\eqref{eq:58} represent the dispersion relations for free waves,
that can spread in the studied system.

\section{Light delay phenomenon for the two-level system in BEC
state}\label{s5}

In the previous section the expression for the permittivity of the
system in BEC state (see Eq.~\eqref{eq:54}) was found. For the
system with the frequency of the external laser field tuned up to
the difference between two defined levels (marked below by indexes 1
and 2, $\omega\approx\Delta\varepsilon_{21}$) it can be written in
more suitable form:
\begin{equation}\label{eq:59}                              %%%% ---- %%%% Eq. 59
    \epsilon^{-1}(\textbf{k},\omega)
    \approx1+\dfrac{g_1 g_2|\sigma_{12}
    (\textbf{k})|^2}{2\pi^2 k^2}
    \dfrac{(\nu_{1}-\nu_{2})}
    {\delta\omega+i\gamma}.
\end{equation}
Here $g_i$ is the degeneracy order of $i$ level,
$\delta\omega=\omega-\Delta\varepsilon_{21}$ is the laser detuning,
$\gamma\equiv\gamma_{12}$ is linewidth related to the transition
probability from the upper to the lower state. Note that we also
neglected the term $\varepsilon_{k}$, the substantiation of such
operation will be discussed below.

To study the propagation properties it is more convenient to turn to
the refractive index and damping factor quantities. To do it we set
the magnetic permeability that enters the dispersion relation
\eqref{eq:58} close to unity $\mu(\textbf{k},\omega)=1$. In that
case the refractive index and damping factor are concerned with the
permittivity by well-known relations:
\begin{equation}\label{eq:60}                              %%%% ---- %%%% Eq. 60
    n=\dfrac{\sqrt{\epsilon'
    +\sqrt{\epsilon'^2+\epsilon''^2}}}
    {\sqrt{2}},\quad
    \varkappa=\dfrac{\sqrt{-\epsilon'
    +\sqrt{\epsilon'^2+\epsilon''^2}}}
    {\sqrt{2}}.
\end{equation}
Here $\epsilon'$ and $\epsilon''$ are the real and imaginary part of
the permittivity:
\begin{equation*}                              %%%% ---- %%%%
    \epsilon(\textbf{k},\omega)
    =\epsilon'(\textbf{k},\omega)
    +i\epsilon''(\textbf{k},\omega),
\end{equation*}
which can be found from the expression \eqref{eq:59} by taking the
real and imaginary part:
\begin{equation}\label{eq:61}                              %%%% ---- %%%% Eq. 61
    \epsilon'=\dfrac{\delta\omega
    (\delta\omega+a)+\gamma^2}
    {(\delta\omega+a)^2+\gamma^2},\quad
    \epsilon''=\dfrac{\gamma a}
    {(\delta\omega+a)^2+\gamma^2},
\end{equation}
where
\begin{equation}\label{eq:62}                              %%%% ---- %%%% Eq. 62
    a(\textbf{k})=(\nu_{1}-\nu_{2})\dfrac{g_1 g_2
    |\sigma_{12}(\textbf{k})|^2}{2\pi^2 k^2}.
\end{equation}

Now one can find the dependence of the group velocity on the system
parameters:
\begin{equation}\label{eq:63}                              %%%% ---- %%%% Eq. 63
    v_{g}=\dfrac{c}
    {n+\omega(\partial n/\partial \omega)}.
\end{equation}
As it easy to see, the ultraslow light phenomenon can be observed in
case when matter has a rather strong dispersion, or, according to
the formula \eqref{eq:63},
\begin{equation}\label{eq:64}                              %%%% ---- %%%% Eq. 64
    \omega(\partial n/\partial \omega)\gg1.
\end{equation}
It is known that to use the group velocity concept the energy
dissipation must be rather small, or, in other words, the next
condition must take place:
\begin{equation}\label{eq:65}                              %%%% ---- %%%% Eq. 65
    |\epsilon'|\ll|\epsilon''|
\end{equation}
If $\epsilon'\sim1$, then according to the expression \eqref{eq:60}
and relation \eqref{eq:65}  $(\partial n/\partial \omega)\approx0.5
(\partial \epsilon'/\partial \omega)$. The partial derivative of the
permittivity real part can be found from the formula \eqref{eq:61}:
\begin{equation}\label{eq:66}                              %%%% ---- %%%% Eq. 66
    \dfrac{\partial \epsilon'}{\partial \omega}
    =\dfrac{a\left[(\delta\omega+a)^2-\gamma^2
    \right]}{\left[(\delta\omega+a)^2+\gamma^2
    \right]^2}
\end{equation}
Using Eqs. \eqref{eq:63}-\eqref{eq:66} for the slowed signal group
velocity finally we get:
\begin{equation}\label{eq:67}                              %%%% ---- %%%% Eq. 67
    v_{g}\approx2c\dfrac
    {\left((\delta\omega+a)^2+\gamma^2
    \right)^2}
    {a\omega\left[(\delta\omega+a)^2-\gamma^2
    \right]}
\end{equation}

Now let us consider the cases in which the light delay phenomenon
for the two-level system in BEC state can be observed. To do it we
shall go to the limit $\delta\omega\rightarrow0$. In that case
according to Eq.~\eqref{eq:61} the real and imaginary parts of
permittivity will have the following limits:
\begin{equation*}                              %%%% ---- %%%%
    \lim\limits_{\delta\omega\rightarrow0}
    \epsilon'=\dfrac{\gamma^2}
    {\gamma^2+a^2},\quad
    \lim\limits_{\delta\omega\rightarrow0}
    \epsilon''=\dfrac{\gamma a}
    {\gamma^2+a^2},
\end{equation*}
Now one can see that the condition for the dissipation smallness
\eqref{eq:65} is equivalent to the following condition:
\begin{equation*}                              %%%% ---- %%%%
    \dfrac{|a|}{\gamma}\ll1,
\end{equation*}
Note, that it is necessary also to add the slowing down (strong
dispersion) condition \eqref{eq:64}, which in this case can be
written as:
\begin{equation*}                              %%%% ---- %%%%
    \dfrac{c}{v_g}\sim\Delta\varepsilon_{21}
    \dfrac{|a|}{\gamma^2}\gg1,
\end{equation*}
As for the defined system with fixed energy spectrum structure the
one parameter that can be changed is occupation difference
$\delta\nu=(\nu_{1}-\nu_{2})$, which is included in the parameter
$a$ (see \eqref{eq:62}), thus, basing on these relations we get the
expression that characterizes the region where the mentioned
phenomenon can be observed:
\begin{equation}\label{eq:68}                              %%%% ---- %%%% Eq. 68
    \dfrac{\gamma}{\Delta\varepsilon_{21}}
    \ll\dfrac{|a|}{\gamma}\ll1,
\end{equation}

Let us demonstrate that such region can exist on the example of
cesium hyperfine structure ground state levels. The choice of
hyperfine levels is stimulated by their stability and pumping
capability (see more in that case in Ref. \cite{ref:03}). Note that
for such levels the dipole transitions are forbidden, thus the
transitions come from the higher order effects that reflects on the
extremely little values of linewidths. It will be shown below that
such fact gives the opportunity for signal to propagate with small
loss of energy.

It should be mentioned that the description analogically can be
spread to the other hydrogen-like atoms and other type of levels.
For example, it can be used for description the experiments with
ultraslow light in BEC of sodium atoms \cite{ref:04}, where for the
pulse slowing the three-level system had been used.

As good known, alkali metals ($^{133}$Cs in particular) in the
ground state do not have the dipole moment, thus the charge density
matrix element $\sigma_{12}$ (see definition \eqref{eq:18}) must be
expanded to the second order over $(\textbf{ky})\ll1$. As a result,
one gets:
\begin{equation}\label{eq:69}                              %%%% ---- %%%% Eq. 69
    \sigma_{12}(k)\approx\dfrac{e}{3}(kr_0)^2,
\end{equation}
where $r_0$ is the atomic radius (for the cesium ground state
$r_0\approx 2,6\times10^{-8}$~cm \cite{ref:05}), $e$ is the electron
charge. Taking $g_1=7$, $g_2=9$, the linewidth
$\gamma\approx3,8\times10^{-21}~\text{eV}$ (or $10^{-6}$~Hz in
frequency units, as in "cold atomic clocks" experiments
\cite{ref:06}), $k\simeq(\Delta\varepsilon_{21}/c)$, where
$\Delta\varepsilon_{21}\approx 3,8\times10^{-5}$~eV (microwaves with
frequency 9,2~GHz) and basing on the expressions \eqref{eq:62},
\eqref{eq:68} one can find the next region for the occupation
difference when light delay phenomenon can be observed:
\begin{equation}\label{eq:70}                              %%%% ---- %%%% Eq. 70
    10^{-3}~\text{cm}^{-3}
    \ll|\nu_1 - \nu_2|\ll
    3\times10^{13}~\text{cm}^{-3}.
\end{equation}
In the experiments for the BEC regime reaching the cesium atoms with
peak density of $\nu=7\times10^{10}~\text{cm}^{-3}$ was kept in the
trap \cite{ref:07}, so, according to the inequality \eqref{eq:70},
one can conclude that also for such experiments the microwave pulse
slowing down phenomenon can be observed. Note that the effect
becomes greater with the density difference increasing until it
reaches the upper limit of the expression \eqref{eq:70}, when
damping effects in the system are great.

Also we must note the following. As it easy to see from
Eq.~\eqref{eq:67} in the limit $\delta\omega\rightarrow0$, the sign
of the group velocity $v_{g}$ depends directly on the sign of the
quantity $a$, that in turn depends on the sign of the difference
$(\nu_1 - \nu_2)$. In other words, it depends whether population is
normal or inverse.

In case of normal population $(\nu_1 > \nu_2)$ the group velocity of
signal is negative. It is traditionally considered that the group
velocity for the transparent matter is positive. But here one can
conclude that due to the relation~\eqref{eq:68}, the signal can
propagate in the system with rather small dissipation
(\textit{i.e.}, in fact, the matter is transparent) and rather slow
velocity. Let us note that an observing of electromagnetic pulses
with negative group velocity is not so abnormal. The existence of
such kind of phenomena for physical systems when the wave frequency
is close to atomic (or molecular) resonances was pointed out and
studied in many works (both theoretical, see                            %% CHANGED !!! %%
\textit{e.g.}~\cite{ref.Kad78}, and
experimental~\cite{ref.Chu82,ref.Macke85}). In case of inverse
population $(\nu_1 < \nu_2)$ more "normal" situation takes place
because the group velocity of the slowed pulse is positive.

We stress that such rather unusual phenomenon occurs due to the
unique property of hyperfine splitted ground state levels
\eqref{eq:69}. One can show that for the two-level system with
allowed dipole transitions the signal will not propagate due to
large absorbtion. The most obvious way out from this situation is
the second (coupling) laser using that leads to the
electromagnetically induced transparency (EIT, see more in that case
in Refs.~\cite{ref:04, ref:08})

Now, let us say a few words about the quantity $\varepsilon_{k}$,
that had been neglected in deriving the equation~\eqref{eq:59}. If
to take $k=(\Delta\varepsilon_{21}/c)$, one can find for cesium
${\varepsilon_{k}\approx3,5\times10^{-30}}$~eV. So, even at the
point $\delta\omega=0$ it is small in comparison with linewidth
$\gamma$, thus, one can note that used approximation is correct.

\section{Conclusion}

Thus, by using the microscopic approach, we studied the linear
response of the system with bound states of particles to disturbing
effect of an external electromagnetic field. Our approach was based
on a novel formulation of the second quantization method in the
presence of bound states of particles \cite{ref:02}. Such approach
allowed to obtain the expressions for the macroscopic
characteristics of the ideal hydrogen-like plasma at low
temperatures taking into account not only the contribution of free
charged fermions but also their bound states (alkali atoms). The
expression for the dielectric permittivity of the ideal gas of
alkali atoms in the presence of Bose-Einstein condensation phase was
also obtained. The dispersion equation for the waves propagating in
the system was derived and the existence of resonance frequencies
was found.

Our approach gave the opportunity to study the propagation
properties of the microwave signal, tuned up to the transition
between two hyperfine ground state levels of alkali atoms that are
considered in BEC state. In contrast to the dipole allowed
transitions, it was shown that the signal could propagate in such
system with rather small energy loss. Such fact allowed to introduce
the group velocity concept. The slowing down conditions for the
signal that propagates in BEC were studied. Moreover, we revealed
the dependence of the group velocity sign on the population
difference sign.

\end{document}